\documentclass[twoside]{article}
\usepackage[a4paper]{geometry}
\usepackage[utf8x]{inputenc}
\usepackage[T1]{fontenc}
\usepackage{lmodern}
\usepackage{fRReE}
\usepackage{url}
\usepackage{hyperref}
\usepackage{natbib} 
\usepackage{graphicx}
\usepackage{float} 
\usepackage{listings}

\RRlogo{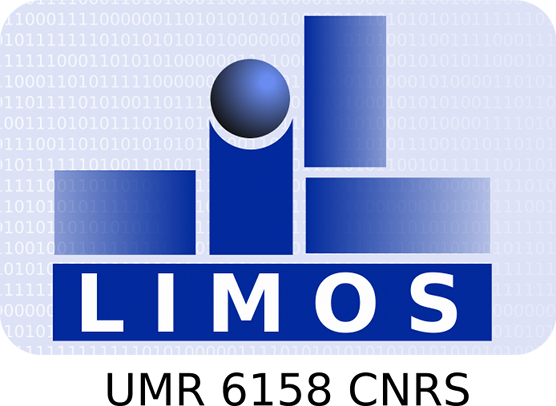}

\RRrc{limos}
\RRrcname{LIMOS - UMR CNRS 6158}
\RRrcaddr{
Campus des C\'ezeaux\\
B\^atiment ISIMA\\
BP 10125 - 63173 Aubi\`ere Cedex\\
France}
\RRrcwebsite{\url{http://limos.isima.fr/}}

\RRdate{October 2011}

\RRauthor{
Jonathan \textsc{Passerat-Palmbach}\RRnote{This author's PhD is partially funded by the Auvergne Regional Council and the FEDER}
\RRaffiliation[sfn0]{ISIMA, Institut Sup\'erieur d'Informatique, de Mod\'elisation et de ses Appplications, BP 10125, F-63173 AUBIERE}
\RRaffiliation[sfn1]{Clermont Universit\'e, Universit\'e Blaise Pascal, LIMOS, BP 10448, F-63000 CLERMONT-FERRAND}
\RRaffiliation[sfn3]{CNRS, UMR 6158, LIMOS, F-63173 AUBIERE}
\and
\\Jonathan \textsc{Caux}
\RRaffiliationref{sfn0}
\RRaffiliationref{sfn1}
\RRaffiliationref{sfn3}
\RRaffiliation[ibc]{Integrative Biocomputing, Nouvelles Structures - Place du Granier, F-35135 CHANTEPIE}
\and
\\Pridi \textsc{Siregar}
\RRaffiliationref{ibc}
\and
\\Claude \textsc{Mazel}
\RRaffiliationref{sfn0}
\RRaffiliationref{sfn1}
\RRaffiliationref{sfn3}
   \and
\\David R.C. \textsc{Hill}
\RRaffiliationref{sfn0}
\RRaffiliationref{sfn1}
\RRaffiliationref{sfn3}
}

\RRauthorhead{Passerat-Palmbach et. al}
\RRtitle{Warp-Level Parallelism: Enabling Multiple Replications In Parallel on GPU}
\RRsource{European Simulation and Modeling Conference 2011}
\RRtitlehead{ESM 2011}

\RRcopyright{\textcopyright 2011 Eurosis}
\RRoriginalpages{76-83}
\RRadditionalinformation{ISBN: 978-90-77381-66-3\\\textit{Best paper award}}

\RRabstract{
\textit{Stochastic simulations need multiple replications in order to build confidence intervals for their results.
 Even if we do not need a large amount of replications, it is a
good practice to speed-up the whole simulation time using the Multiple
Replications In Parallel (MRIP) approach. This approach usually
supposes to have access to a parallel computer such as a symmetric
multiprocessing machine (with many cores), a computing cluster or a
computing grid. In this paper, we propose Warp-Level Parallelism (WLP), a GP-GPU-enabled
solution to compute MRIP on GP-GPUs (General-Purpose Graphics
Processing Units). These devices display a great amount of
parallel computational power at low cost, but are tuned to process efficiently the same operation on several data, through different threads. Indeed, this paradigm is called Single Instruction, Multiple Threads (SIMT). Our approach proposes to rely on small threads groups, called warps, to perform independent computations such as replications. We have benchmarked WLP with three
different models: it allows MRIP to be computed up to six times faster than
with the SIMT computing paradigm.}

}
\RRkeyword{Stochastic Simulation; Multiple Replications In Parallel (MRIP); GP-GPU; CUDA; Warp-Level Parallelism (WLP)}

\begin{document}
\makeRR

\section{INTRODUCTION}

Replications are a widespread method to obtain confidence intervals for
stochastic simulation results. It consists in running the same
stochastic simulation with different random sources and averaging the
results. According to the Central Limit Theorem, the average result is approximated in an accurate enough way by a Gaussian Law, for a number of replications greater than 30. Thus, for a number of replications greater than 30, we can obtain a confidence interval with a satisfactory precision.

\bigskip

There are many cases where a single simulation can last for a while, so
30 of them run sequentially may represent a very long computation time.
Because of this overhead, 30 replications are hardly run in most
simulations. Instead, a good practice is often to run 3 replications
when debugging, and 10 replications are commonly used to compute a
confidence interval. To maintain an acceptable computation time while
running 30 or more replications, many scientists proposed to run in
parallel these independent simulations. This approach has been named
Multiple Replication in Parallel (MRIP) in the nineties
\cite{Pawlikowski.etal.1994}. As its name suggests, its main idea is to run
each replication in parallel \cite{Hill1997, Pawlikowski2003}. In addition,
when we explore an experimental plan we have to run different sets of
replications, with different factor levels according to the
experimental framework \cite{Hill1996, Amblard.etal.2003}. In this paper, we
will not consider any constraints that need to be satisfied when
implementing MRIP. One of the main barriers that often prevents
simulationists to achieve a decent amount of replications is, on the
one hand, the lack of knowledge in the parallelization techniques and
on the other hand the parallel computing facilities available. Our work
tackles this problem by introducing a way to harness the computational
power of GP-GPUs (General-Purpose Graphics Processing Units – GPUs
hereafter), which are rather cheap compared to regular parallel
computers, to process MRIP quicker than on a scalar CPU (Central
Processing Unit).

\bigskip

GPUs deliver such an overwhelming power at a low cost that they now play
an important role in the High Performance Computing world. However,
this kind of devices display major constraints, tied to its intrinsic
architecture. Basically, GPUs have been designed to deal with
computation intensive applications such as image processing. One of
their well-known limits is memory access. Indeed, since GPUs are
designed to be efficient at computation, they badly cope with
applications frequently accessing memory. Except by choosing the right
applications, the only thing we can do to overcome this drawback is to
wait for the hardware to evolve in such a way. Last NVIDIA GPUs
generations, codenamed Fermi, show a move in this way by improving
cache memories available on the GPU. This leads to better performances
for most applications at no development cost, only by replacing the old
hardware by the state-of-the-art one.

\bigskip

Now, what we can actually think about is the way we program GPUs.
Whatever the programming language or architecture one chooses to
develop his application with, CUDA (Compute Unified Device
Architecture) or OpenCL, the underlying paradigm is the same: SIMT
(Single Instruction, Multiple Threads). Thus, applications are tuned to
exploit the hardware configuration, which is a particular kind of SIMD
architecture (Single Instruction, Multiple Data). To obtain speed-ups,
we must propose parallel applications that will be SIMD compliant. This
point reduces the scope of GPU-enabled applications.

\bigskip

In the SIMT paradigm, threads are automatically grouped into 32-wide
bundles called warps. Warps are the base unit used to schedule both
computation on Arithmetic and Logic Units (ALUs) and memory accesses.
Threads within the same warp follow the SIMD pattern, i.e. they are
supposed to execute the same operation at a given clock cycle. If they
do not, a different execution branch is created and executed
sequentially every time a thread needs to compute differently from its
neighbours. The latter phenomenon is called branch divergence, and leads to
significant performance drops. However, threads contained in different
warps do not suffer the same constraint. They are executed
independently, since they belong to different warps.

\bigskip

In this paper, we introduce Warp-Level Parallelism (WLP), a paradigm to
evaluate the approach of using GPUs to compute MRIP, using an
independent warp for each replication. Our study will:

\begin{itemize}
\item Describe a mechanism to run MRIP on GPU;
\item Propose an implementation of our approach: WLP;
\item Benchmark WLP with three different simulation models.
\end{itemize}

\section{GENERAL CONCEPTS OF GPU PROGRAMMING AND ARCHITECTURE}
This section does a brief recall of the major concepts introduced by GPU
programming and especially by CUDA. It also basically describes how a
GPU architecture is organized, since these aspects are directly tied to
our approach.

\subsection{The Single Instruction Multiple Threads (SIMT) paradigm}
SIMT is the underlying paradigm of any CUDA application. It is based on
the well-known SIMD paradigm. While using SIMD, the same instruction is
executed in parallel on multiple computational units, but take
different data flows in input. Instead of viewing SIMT as a simple SIMD
variant, one needs to understand that it has been created to simplify
applications development on GPU. The main idea is first to allow
developers to deal with a unique function, named a kernel, which is
going to be run in parallel on the GPU. Second,
developers manipulate threads in SIMT, which are a much more common tool nowadays than traditional vectors enabling SIMD
parallelization.

\bigskip

In order to handle SIMT more easily, CUDA introduces different bundles of threads. As a matter of fact, threads are grouped
into blocks, which size and 3D-geometry are defined by the user. The
whole blocks of a kernel form a 2D grid. Each thread will be uniquely
identified in the kernel thanks to an identifier computed from a
combination of its own coordinates and of its belonging
block's. More precisely, in addition to grid and
blocks, CUDA devices automatically split threads into fixed-size
bundles called warps. Currently, warps contain 32 threads. This group
is extremely important in the low-level mechanisms running on a GPU.

\bigskip

As long as NVIDIA has defined both its GPU architecture and the SIMT
paradigm, the latter is not only convenient, it also perfectly fits its
host architecture. Its sole purpose is to be used on GPU architecture,
which is quite different from other multi-core architectures, especially
from CPU ones, as we will see in the next part.

\subsection{Basic architecture of a GPU}
While a CPU possesses few cores, each of them allowing the execution of
one thread at a time, a GPU possesses a small number of Streaming
Multiprocessors (SM) (for instance an NVIDIA Fermi C2050 has 14 SMs).
Each SM embeds an important number of computational units (there are 32
floating point computational units - called Streaming
Processor (SP) - on each SM of a Fermi C2050). In theory,
the floating-point computation power of a GPU board is equal to the
number of SMs multiplied by the number of SPs. Another figure that
needs to be considered in the architecture is the number of warp
schedulers. The latter are key elements of CUDA performance. In fact,
memory accesses are done per warp. However, because of memory latency,
the warps-schedulers select the warps that have their data ready to
process. Consequently, the more warps can be scheduled, the more the
memory latency can be hidden.

\bigskip

\begin{figure}[H]
\centering
\includegraphics[keepaspectratio, scale = 0.64]{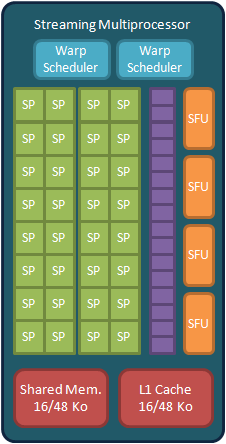} 
\caption{Simplified architecture of an NVIDIA Fermi C2050 streaming multiprocessor}
\label{architectureFermi}
\end{figure}

When the former generation of NVIDIA GPUs was issued with a single
warp-scheduler per SM \cite{Lindholm.etal.2008}, Fermi now owns two warp
schedulers per SM \cite{Wittenbrink.etal.2011}. They are first employed when threads need to be scheduled on the SM
they have been assigned to. In fact, threads within a warp also achieve memory
accesses in parallel, before processing the same instruction on these
data. To sum up, when threads are bound to each other, and must execute
the same instructions according to SIMD machinery, warps are the
smallest unit that run in parallel on the different SMs of a GPU, and
are the smallest GPU element that is able to process independent code
sections. Indeed, given that different warps either run on different
SMs, or on the same but at different clock ticks, they are fully
independent to each other. Figure \ref{architectureFermi} shows a simplified
representation of a SM of the Fermi architecture.

\subsection{Blocks dispatching and warps scheduling in NVIDIA GPUs}
Now that we have introduced the basic functioning of CUDA-enabled GPUs,
let us detail the particular features that will help us to achieve MRIP
on such architectures. We will see in this part how our GPU-enabled
MRIP implementation relies on the scheduling features provided by
NVIDIA CUDA devices. Previously, NVIDIA GPUs were only able to run a
single kernel at a time. Thus, blocks of threads were dispatched
through all the available SMs in a more or less logical way: SMs were
activated in turn, striding indices four by four. When every SM had
been activated, the process started again.

\bigskip

One of the key features of the cutting-edge Fermi architecture is the
ability to run several kernels in parallel on the same device. To do
so, the way blocks of threads are dispatched through the device has
been redesigned in a new fashion. Now, every block of threads, no
matter which kernel it belongs to, is first handled by a top-level
scheduler referred to as the GigaThread Engine. It is supposed to
dispatch blocks of threads to the Streaming Multiprocessors (SMs). The
point is CUDA has always proposed asynchronous kernels calls to
developers. Now that Fermi-enabled devices can run several kernels in
parallel, GigaThread needs to take into account any potential upcoming
kernel. Consequently, the dispatcher cannot reserve all the SMs to run
a first kernel, given that a second one could be launched at any time.
When the second kernel appears, some resources will still be available
so that they can be assigned to the new kernel blocks.

\bigskip

Moreover, GigaThread enables immediate replacement of blocks on an SM
when one completes executing. Since context switching has been fastened
with Fermi, blocks of threads can fully take advantage of the hardware
device thanks to GigaThread dispatching capabilities. From an external
point of view, and since we do not have the real specifications, we
have noted that the dispatching of blocks does not seem to be
deterministic. NVIDIA uses a specific way to place blocks on SMs:
indeed, SMs will not be enabled in order. SMs bearing non-consecutive
identifiers will in fact run consecutively ordered blocks.

\section{A WARP MECHANISM TO SPEED UP REPLICATIONS}
Two problems arise when trying to port replications to GPU threads,
considering a replication per thread. First, we generally compute few
replications, whereas we have seen that GPUs needed to achieve large
amounts of computations to hide their memory latency. Second,
replications of stochastic simulations are not renowned for their
SIMD-friendly behaviour. Usually, replications fed with different
random sources will draw different random numbers at the same point of
the execution. If a condition result is based on this draw, divergent
execution paths are likely to appear, forcing threads within a same
warp to be executed sequentially because of the intrinsic properties of
the device.

\bigskip

The idea that we propose in this paper is to take advantage of the
previously introduced warp mechanism to enable fast replications of a
simulation. Instead of having to deal with Thread-Level Parallelism
(TLP) and its constraints mentioned above, we place ourselves at a
slightly higher scope to manipulate warps only. Let this paradigm be
called Warp-Level Parallelism (WLP), as opposed to TLP. Now running
only one replication per warp, it is possible to have each replication
to execute different instructions without being faced to the branch
divergence problem.

\bigskip

But to successfully enable easy development of simulation replications
on GPU using one thread per warp, two mechanisms are needed. 

\begin{figure}[!h]
\centering
\includegraphics[keepaspectratio, scale = 0.40]{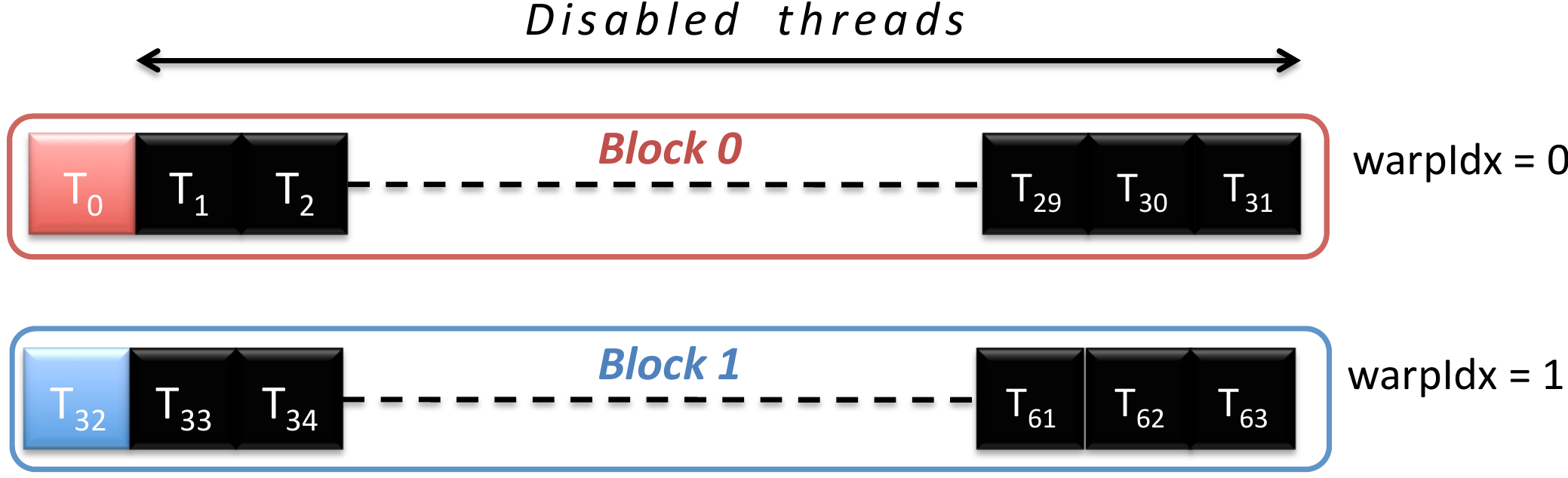} 
\caption{Representation of thread disabling to place the application at a warp-level}
\label{disabled_threads}
\end{figure}

\bigskip
First, it is necessary to restrict each warp to use only one valid thread. By
doing so, we ensure not to have divergent paths within a warp. Moreover,
we artificially increase the device’s occupancy, and consequently, we
take advantage of the quick context switching between warps to hide
slow memory accesses. Theoretically, we should use the lowest block size maximizing occupancy. For instance, 
a C2050 board owns 14 SMs, and can schedule at most 8 blocks per SM. In this case, the optimal block size when running 50 replications would be 32 threads per block. This situation is represented in Figure \ref{disabled_threads}, where we can see two warps
running their respective first threads only. The 31 remaining threads
are disabled, and will stall until the end of the kernel.
Unfortunately, the GigaThread scheduler, introduced in the previous section, does not always enable a kernel to run on every available SM.
In addition, SMs' memory constraints might compromise this ideal case by reducing the number of available blocks per SM.

Second, there has to be an easy solution to get a unique index for each
warp. TLP relies essentially on threads identifiers to retrieve or write
data back. Thus, WLP needs to propose an equivalent mechanism so that
warps can be distinguished to access and compute their own data.

\bigskip

Thanks to the two tools introduced in this section, it is possible to create a kernel where only one
thread per warp will be valid, and where it will be easy to make each
valid thread compute different instructions, or work on different data
depending on the new index.

\bigskip

Although we could not figure out the real behavior of the GigaThread
Engine dispatcher, the characteristics
noticed in this part are sufficient to evaluate the performance of the dispatching policy. 
Furthermore, the new scheduling features introduced in
Fermi significantly enhance the overall performance of our warp-based
approach, given that it highly relies on warp scheduling and block
dispatching.

\section{IMPLEMENTATION}
Now that we have defined our solution, we will propose an implementation
in this section. To do so, we need to focus on two major constraints:
first, we should keep a syntax close to C++ and CUDA, so that users are
not confused when they use our approach. Second, we need to propose
compile-time mechanisms as much as possible. Indeed, since WLP only exploits a restricted amount of the
device's processing units, we have tried to avoid any
overhead implied by our paradigm.

\bigskip

This paper intends to prove that our approach is up and running. Thus, this
section will only introduce a restricted number of keywords used by
WLP. As we have seen previously, we first have to be able to identify
the different warps, in the same way SIMT does with threads. One way to obtain the warp identifier is to compute it at
runtime. Indeed, we know that warps are formed by 32 threads in current
architectures [NVIDIA2011a]. Thus, knowing the running kernel
configuration thanks to CUDA defined data-structures, we are able to
figure out the warp identifier with simple operations only, similarly to what have
done \cite{Hong.etal.2011}. The definition of a \texttt{warpIdx} variable
containing the warp's identifier can be written as in
Figure \ref{const_warpidx}:

\begin{figure}[H]
\centering 
\begin{lstlisting}
const unsigned int warpIdx = ( 
   threadIdx.x + blockDim.x * (
	threadIdx.y + blockDim.y * (
		threadIdx.z + blockDim.z * (
			blockIdx.x + gridDim.x * blockIdx.y
   ) ) ) ) / warpSize;
\end{lstlisting}

\caption{Const-definition of warpIdx}
\label{const_warpidx}
\end{figure}

Conceptually, this definition is ideal because \texttt{warpIdx} is declared as a
‘constant variable’, and the warp identifier does not change during a
kernel execution. This formula fits with the CUDA way to number threads, which first considers threads' x indices, then y and finally z, within a block. The same organization is applied to blocks numbering \cite{Kirk.Hwu2010}.
Please note that the \texttt{warpSize} variable is provided by CUDA. This makes our implementation portable since warp sizes may evolve in future
CUDA architectures.

\bigskip

Although this method introduces superfluous computations to figure out the kernel's configuration, we find it easier to understand for developers.
Another way to compute the warp's identifier would have been to write CUDA PTX assembly\cite{NVIDIA2011}. The latter is the Instruction Set Architecture (ISA)
currently used by CUDA-enabled GPUs. CUDA enables developers to insert
inlined PTX assembly into CUDA high-level code, as explained in
[NVIDIA2011b]. However, this method is far less readable than ours, and would not be more efficient since we only compute \texttt{warpIdx} once: at initialization.
\bigskip

This warp identifier will serve as a base in WLP. When
classical CUDA parallelism makes a heavy use of the runtime-computed
global thread identifier, WLP proposes \texttt{warpIdx} as an equivalent.

\bigskip

Now that we are able to figure out threads' parent warps,
let us restrain the execution of the kernel to a warp scope. Given that
we need to determine whether or not the current thread is the first
within its belonging warp, we will be faced to problems similar to
those encountered when trying to determine the warp identifier. In
fact, a straightforward solution reckoning on our knowledge of the
architecture quickly appears. It consists in determining the global
thread identifier within the block to ensure it is a multiple of
the current warp size. Once again, the kernel configuration is issued
by CUDA intrinsic data structures, but we still need a reliable way to
get the warp size to take into account any potential evolution.
Luckily, we can figure out this size at runtime thanks to the aforementioned \texttt{warpSize} variable. Consequently, here is how we begin a warp-scope kernel in WLP:

\begin{figure}[H]
\centering 
\begin{lstlisting}
if ( ( threadIdx.x + blockDim.x * 
    ( threadIdx.y + blockDim.y * threadIdx.z ) )
\end{lstlisting}

\caption{Directive enabling warp-scope execution}
\label{warpsz_asm}
\end{figure}

We now own the bricks to perform WLP, but still lack a user-friendly
API. Indeed, it would not be adapted to ask our users to directly use
complex formulas without having wrapped them up before in higher-level
calls. To do so, we chose to use macros, for the sake that they are
compile-time mechanisms, thus not causing any runtime overhead, and
that they are perfectly handled by \textit{nvcc}, the CUDA compiler. Our previous
investigations result in two distinct macros: \texttt{WARP\_BEGIN} and
\texttt{WARP\_INIT}, which respectively mark the beginning of the warp-scope
code portion, and correctly fill the warp identifier variable. When
\texttt{WARP\_INIT} presents no particularities, except the requirement to be
called before any operations bringing into play \texttt{warpIdx}, \texttt{WARP\_BEGIN}
voluntary forgets the block-starting brace following the if
statement. By doing so, we expect users to place both opening and
closing braces of their WLP code if needed, just as they would do with
any other block-initiating keyword.

\bigskip

To sum up, please note once again that this implementation mainly
targets to validate our approach. Still, it lays the foundation of a
more complete API dedicated to WLP. The efficient but not appealing
intrinsic mechanisms are totally masked to users thanks to macros introduced
in WLP.

\section{RESULTS}
In this part, we introduce three well-known stochastic simulation models
in order to benchmark our solution. We have compared
WLP's performances on a Tesla C2050 board to those of
a state-of-the-art scalar CPU: an Intel Westemere running at 2.527 GHz.
For all of the three following models, each replication runs in a
different warp when considering the GPU, whereas the CPU runs the
replications sequentially. The following implementations use L’Ecuyer’s
Tausworthe three-component PRNG, which is available on both CPU and GPU
respectively through Boost.Random and Thrust.Random \cite{Hoberock.Bell2010} libraries. Random
streams issued from this PRNG are then split into several
sub-sequences according to the Random Spacing distribution technique \cite{Hill2010}.

\subsection{Description of the models}
First, we have a classical Monte Carlo simulation used to approximate
the value of Pi. The application draws a succession of random points
coordinates. The number of random points present in the quarter of a
unit circle are counted and stored. At the end of the simulation, the
Pi approximation corresponds to a ratio of the points in the quarter of
a unit circle to the total number of drawn points. The output of the
simulation is therefore an approximate of Pi value. This model takes
two input parameters: the number of random points to draw and the
number of replications to compute.

\bigskip

The second simulation is a M/M/1 queue. For each client, the time
duration before its arrival and the service time is randomly drawn. All
other statistics are computed from these values. The program outputs
are the average idle time, the average time in queue of the clients and
the average time spent by the clients in the system. Because it did not
impact the performances, the parameters of the random distribution are
static in our implementation. Only the number of clients in the system
and the number of replications, which modify the execution time, can be
specified when running the application.

\bigskip

The last simulation is an adaptation of the random walk tests for PNRGs
exposed in \cite{Vattulainen.Ala-Nissila1995}. The idea is to simulate a
walker moving randomly on a chessboard-like map. The original application
tests the independence of multiple flows of the same PRNG. To achieve
this, multiple random walkers are run with different initializations of
a generator on identically configured maps. Basically, each walker computes a replication. In the end, we count the number of walkers in every area
of the map. Depending on the PRNG quality, we should find an equivalent
number of walkers in each area. When the original version splits the
map in four quarters, our implementation uses 30 chunks to put the
light on the opportunity of our approach when there are many divergent
branches in an application.

\subsection{Comparison CPU versus GPU warp}
\begin{figure}[!h]
\centering
\includegraphics[keepaspectratio, scale = 0.65]{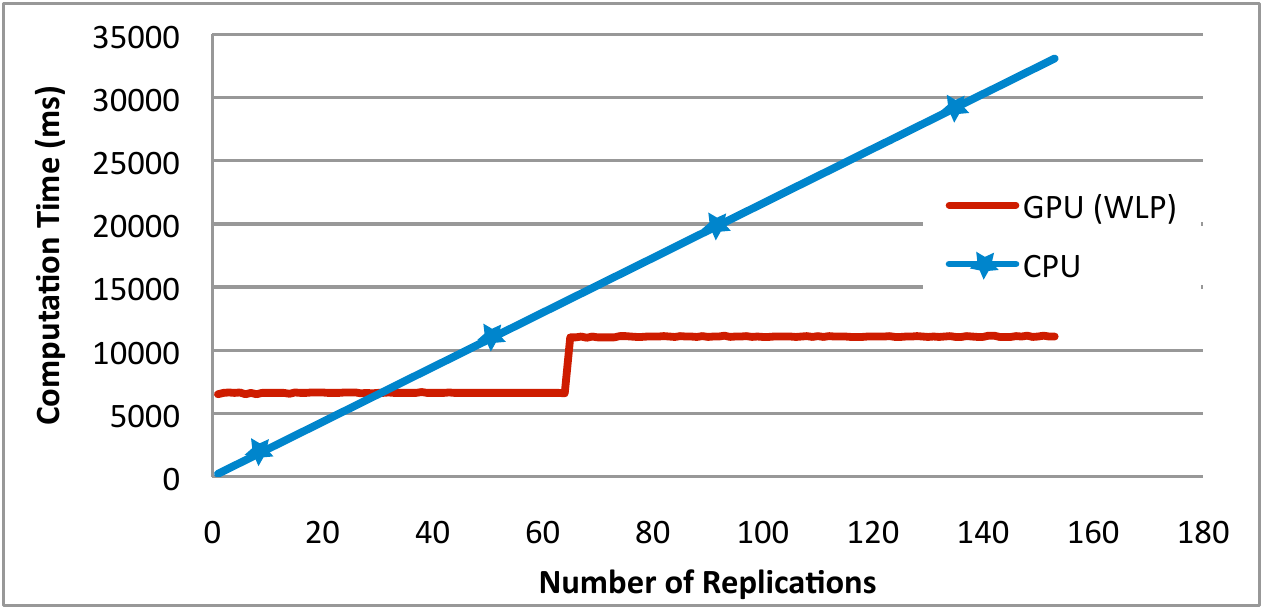} 
\caption{Computation time versus number of replications for the Monte Carlo Pi approximation with 10000000 draws}
\label{mc_pi}
\end{figure}

As we can see in Figure \ref{mc_pi}, the CPU computation time of the
Monte Carlo application approximating the value of Pi grows linearly with the number of replications.
The GPU computation time increases only by steps. This behaviour is due
to the huge parallel capability of the device. Until the GPU card is
fully used, adding another replication does not impact the computation time, because they are all done in parallel. So, when the board is full, any new iteration will increase the computation time. This only happens on the 65$^{th}$ replication because the GPU saved some
resources in case a new kernel would have to be computed simultaneously.
The same mechanism explains that after this first overhead, a new
threshold appears and so on.

\bigskip

Due to this behaviour, GPUs are less efficient than CPUs when the board is
nearly empty. When less than 30 replications are used, more than
two-thirds of the board computational power is idle. Because sequential
computation on CPU is widely faster than sequential computation on GPU,
if only a little of the parallel capability of the card is used, the GPU
runs slower. But when the application uses more of the card parallel
computation power, the GPU becomes more efficient than the CPU.

\begin{figure}[!h]
\centering
\includegraphics[keepaspectratio, scale = 0.65]{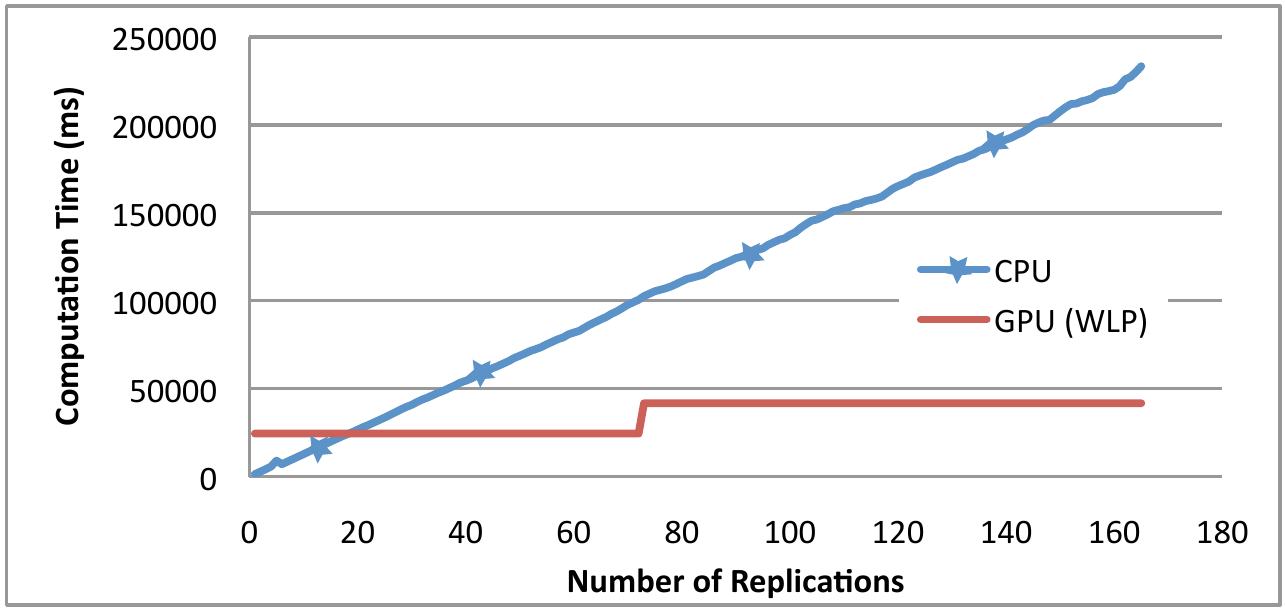} 
\caption{Computation time versus number of replications for a M/M/1 queue model with 10000 clients}
\label{mm1}
\end{figure}

The pattern is very similar for the second model: the M/M/1 queue
(see Figure \ref{mm1}). When the board does not run enough warps
in parallel, the CPU computation is faster than the GPU one. But with
this model, the number of replications needed for the GPU approach to
outperform the CPU is smaller than what we obtained with the previous
simple model. The GPU computation is here faster as soon as 20 replications
are performed, when it required 30 replications to show its efficiency with the first model. This can be
explained by GPUs' architecture, where memory accesses are far more costly than floating point operations in terms of processing time. If the application has a better computational
operations per memory accesses ratio, it will run more efficiently
on GPU. Thus, the GPU approach will catch with the CPU one faster.

\bigskip

This point is very important because it means that depending on the
application characteristics, it can be adequate to use this approach from a certain number of replications, or not. A solution is to
consider the warp approach only when the number of replications is big
enough to guaranty that most of the applications will run faster.

\subsection{Comparison GPU warp versus GPU thread}
If the advantages of WLP-enabled replications compared to CPU ones in
terms of computation time have been demonstrated with the previous
examples, it is necessary to determine if WLP outperforms the classic
TLP.

\bigskip

\begin{figure}[!h]
\centering
\addtolength{\leftskip}{-0.2cm}
\includegraphics[keepaspectratio, scale = 0.64]{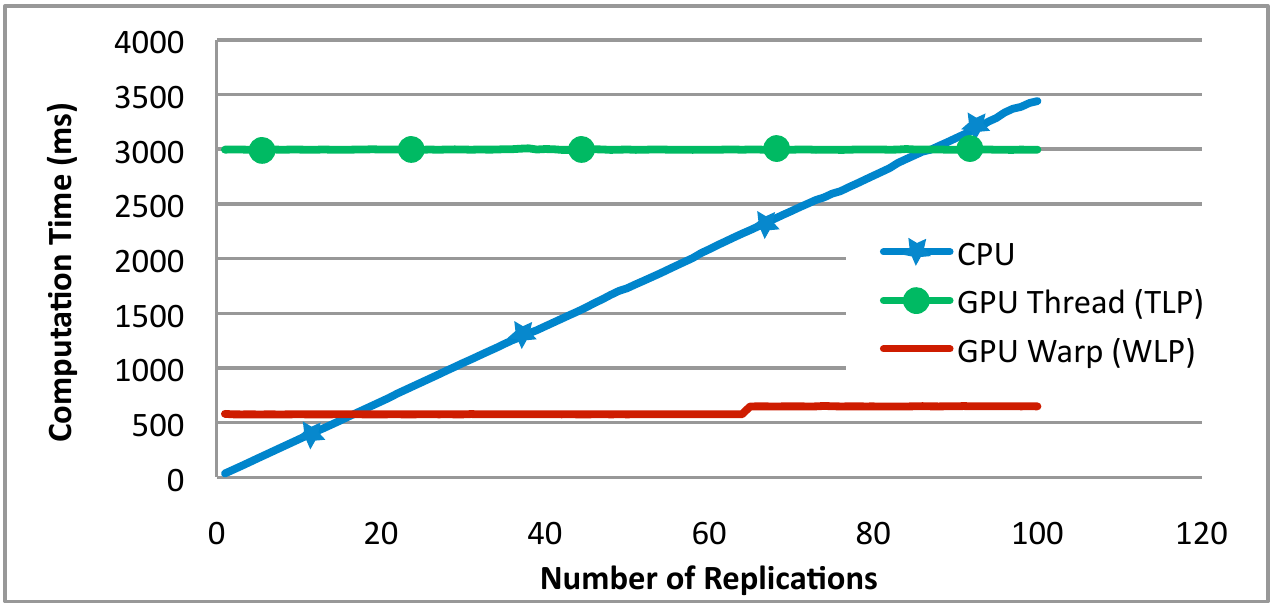} 
\includegraphics[keepaspectratio, scale = 0.64]{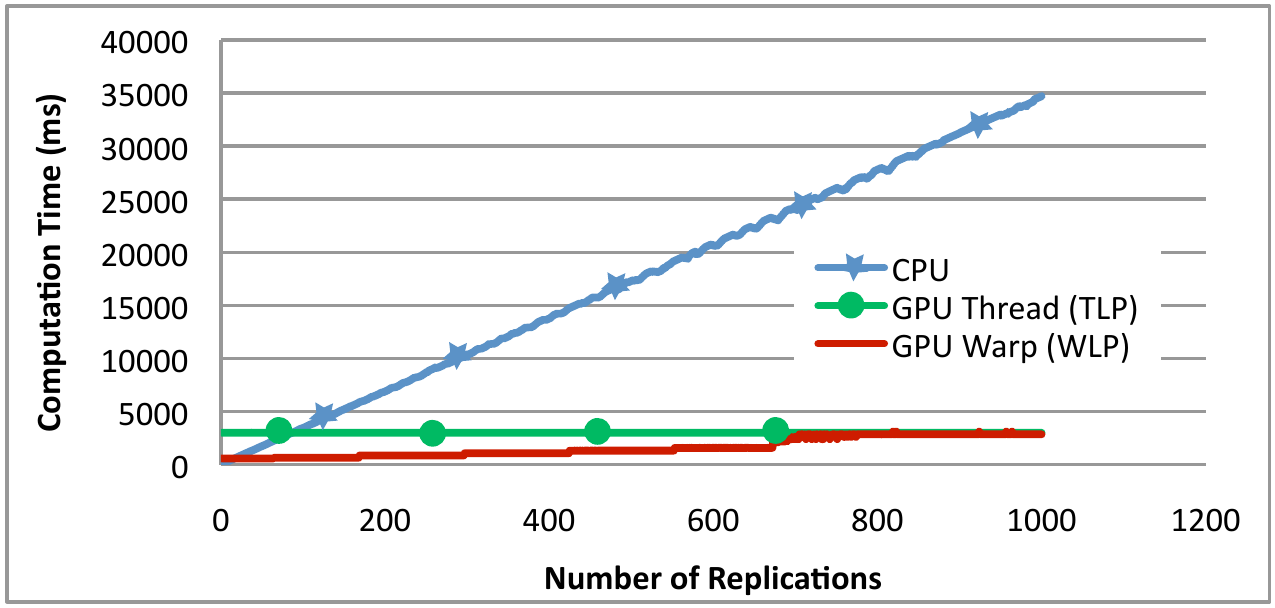} 
\caption{Computation time versus number of replications for a random walk model with 1000 steps \textit{(above: 100 replications, below: 1000 replications)}}
\label{random_walk}
\end{figure}

This case study has been achieved using the last model introduced: our
adaptation of the random walk. Figure \ref{random_walk}$\,$ shows the
computation time noticed for each approach: CPU, GPU with WLP and GPU
with TLP (named \textit{thread} in the caption). Obviously, CPU and
WLP results confirm the previous pattern: the CPU computation time
increases linearly when the WLP one increases by steps. TLP follows logically the same evolution shape as WLP.
Although it is impossible to see it here because the number of
replications is too small, it also evolves step by step, similarly to
the warp approach. WLP consumes a whole warp for each replication. In
the same time, TLP activates 32 threads per warp. Thus, the
latter's steps will be 32 times as long as
WLP's. Having said that, we easily conclude that the
first step in TLP will occur after the 2048$^{th}$ replication.

\bigskip

As we can see in Figure \ref{random_walk}, the computation time needed by
the thread approach is significantly more important than the
computation time of the warp approach (about 6 times bigger for the first 64
replications). But WLP catches
up with TLP when the number of replications increases. When more than
700 replications are performed, the benefit of using the warp approach
is greatly reduced. The best use of the warp approach for this model is
obtained when running between 20 and 700 replications.
Please note that this perfectly matches our replications amount
requirement. It even allows the user to run another set of replications
according to an experimental plan, or to run another set of
replications with a different high quality PRNG. The latter practice is
a good way to ensure that the input pseudo-random streams do not bias
the results. 

\begin{figure}[!h]
\centering
\includegraphics[keepaspectratio, scale = 0.70]{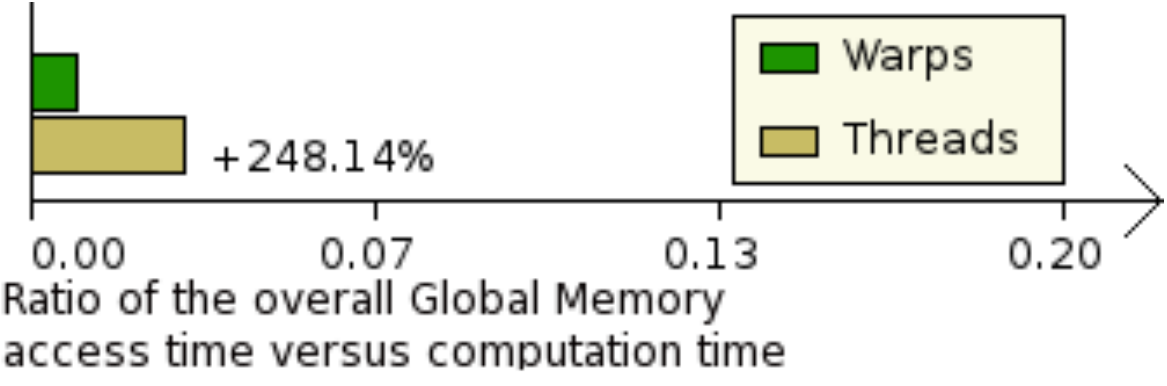} 
\caption{Comparison of TLP and WLP ratio of the overall Global Memory access time versus computation time}
\label{profiler}
\end{figure}

These results are backed up by the output of the NVIDIA Compute Profiler for CUDA applications. The latter tool allows developers to visualize many data about their applications. In our case, we have studied the ratio between the time spent accessing global memory versus computing data. Such figures are displayed in Figure \ref{profiler} for both TLP and WLP versions of the random walk simulation. Our approach obviously outperforms TLP, given that the ratio of overall Global Memory access time versus computation time is about 2.5 times bigger for TLP.

\bigskip

To explain this ratio, let us recall that computation time was lower for WLP. Since the same algorithm is computed by the two different approaches, we should have noticed the same amount of Global Memory accesses in the two cases. In the same way, the profiler indicates significant differences between Global Memory reads and writes for TLP and WLP. These figures are summed up in Table \ref{tableau_read_writes}:

\begin{table}[!h] 
\centering

\begin{tabular}{|c|c|c|}
\cline{2-3}
  \multicolumn{1}{c}{} & 
  \multicolumn{1}{|c}{\textbf{\textit{TLP}}} &
  \multicolumn{1}{|c|}{\textbf{\textit{WLP}}} \\
\hline Reads & 225  & 18 \\
\hline Writes & 302 & 104 \\
\hline 
\end{tabular} 
 \caption{Number of read and write accesses to Global Memory for TLP and WLP versions of the Random Walk} 
 \label{tableau_read_writes}
\end{table}

\bigskip

\section{CONCLUSION}
This paper has shown that using GPUs to compute MRIP was both possible
and relevant. Having depicted nowadays GPUs'
architecture, we have detailed how warp scheduling was achieved on such
devices, and especially how we could take advantage of this feature to
process codes with a high rate of branch divergent parts. Our approach,
WLP (Warp-Level Parallelism), intends to allow users to easily
distribute their experimental plans with replications on GPU.

\bigskip

WLP has been implemented thanks to simple arithmetic operations.
Consequently, WLP displays a minimalist impact on the overall runtime
performance. For the sake of user-friendliness, the internal mechanisms
enabling WLP have been wrapped in high-level macros. At the time of
writing, our version is functional and allows users to create blocks of
code that will be executed independently on the GPU. Each warp will run
an independent replication of the same simulation, determined by the
warp identifier figured out at runtime. By doing so, we prevent
performances to drop as they would do in an SIMT environment confronted
to branch-divergent execution paths. WLP also tackles the GPU underutilization problem by artificially increasing the occupancy.

\bigskip

To demonstrate our approach performances, we have
compared the execution times of a sequence of independent replications
for three different stochastic simulations. Results show that WLP is at
least twice as fast as cutting-edge CPUs when asked to compute a
reasonable amount of replications, that is to say more than 30
replications. This will always be the case when a stochastic simulation
is studied with a design of experiments, where for each combination of
deterministic factors we have to run at least 30 replications, according to the previously mentioned Central Limit Theorem. WLP also
overcomes the traditional CUDA SIMT performances by up to 6 to compute the same
set of replications. Here, SIMT suffers of an underutilized GPU,
whereas WLP takes advantage of a quick warp scheduling.

\bigskip

Insofar performances of WLP increase with the recent Fermi architecture
compared to Tesla, we can expect this approach to be even more
efficient with future CUDA architectures. We will validate this
approach with bigger simulation models. As a matter of fact, two
parameters need to be considered to determine how WLP will scale. On
the one hand, a bigger model will often be more complicated, and will
consequently contain much more divergent branches. When our approach
should benefit of this aspect, on the other hand, bigger models will
also consume more memory, which is the bottleneck of GPU devices.

\bigskip

The current version of WLP forces users to distribute their replications
with our keywords. The target audience of our approach should, for the moment, be
familiar with CUDA or GPU development. To lower the
level of technical difficulty, we are currently thinking about an
automatic tool, taking a simulation model and the number of replications to process in input, and
producing the WLP equivalent in output, thus fully automating MRIP on
GPU.

\bibliographystyle{apalike}
\bibliography{esm2011}

\end{document}